\documentclass[preprint]{revtex4}

\begin{document}

\title{Stochastic Dynamical Structure (SDS) of Nonequilibrium Processes in the Absence of Detailed Balance. III:  \\ potential function in local stochastic dynamics and in steady state of Boltzmann-Gibbs type distribution function }

\author{ L. Yin$^{*}$ and P. Ao$^{\dag}$ }

\address{ $^{*}$School of Physics, Peking University, Beijing 100871, PR China \\
          $^{\dag}$Department of Mechanical Engineering,
               University of Washington, Seattle, WA 98195, USA }

\date{ April 2, 2008}

\begin{abstract}
From a logic point of view this is the third in the series to solve the problem of absence of detailed balance. This paper will be denoted as SDS III.
The existence of a dynamical potential with both local and global
meanings in general nonequilibrium processes has been
controversial. Following an earlier explicit construction
by one of us (Ao, J. Phys. {\bf A37}, L25 '04, cond-mat/0803.4356, referred to as SDS II), in the present paper we show rigorously its existence for a generic class of situations in physical and
biological sciences. The local dynamical meaning of this potential
function is demonstrated via a special stochastic differential
equation and its global steady-state meaning via a novel and
explicit form of Fokker-Planck equation, the zero mass limit. We also give a procedure
to obtain the special stochastic differential equation for any
given Fokker-Planck equation. No detailed balance condition is
required in our demonstration. For the first time we obtain here a
formula to describe the noise induced shift in drift force
comparing to the steady state distribution,
a phenomenon extensively observed in numerical studies. The comparison to two well known stochastic integration methods, Ito and Stratonovich, are made ready. Such comparison was made elsewhere (Ao, Phys. Life Rev. {\bf 2} (2005) 117. q-bio/0605020).    \\
(Please cited present paper as {\it Existence and Construction of Dynamical Potential in Nonequilibrium Processes without Detailed Balance}, L. Yin and P. Ao,   J. Phys. {\bf A39} (2006) 8593-8601.
      http://www.iop.org/EJ/abstract/0305-4470/39/27/003 )
 { } 

\end{abstract}


\maketitle

\section{Formulation of the questions}

A large class of nonequilibrium processes can be described by the
following stochastic differential equation
\cite{langer,prigogine,vankampen,gardiner}:
\begin{equation}  \label{standard}
  \dot{{\bf q}} = {\bf f}({\bf q}) + N_I({\bf q}) \xi(t) \; ,
\end{equation}
where ${\bf f}$ and ${\bf q}$ are $n$-dimensional vectors and
${\bf f}$ a nonlinear function of ${\bf q}$. The noise ${\bf \xi}$
is a standard Gaussian white noise with $l$ independent
components: $\langle \xi_i \rangle = 0$, $\langle \xi_i(t) \xi_j
(t')\rangle=\delta_{ij} \delta (t-t')$, and $i,j=1, 2, ..., l$.
Even in situations that Eq.(\ref{standard}) is not an exact
description, it may still serve as the first approximation for
further modelling \cite{vankampen,gardiner}.
%

A further description of the noise in Eq.(\ref{standard}) is
through the $n\times n$ diffusion matrix $D({\bf q})$, which is
defined by the following matrix equation
\begin{equation}
  N_I({\bf q}) N_I^\tau ({\bf q}) = 2 \epsilon \; D({\bf q}) \; ,
\end{equation}
where $N_I$ is an $n\times l$ matrix,  $N_I^\tau$ is its the
transpose, and $\epsilon$ is a nonnegative numerical constant
playing the role of temperature. This relation suggests that the
$n\times n$ diffusion matrix $D$ is both symmetric and
nonnegative. For the dynamics of state vector ${\bf q}$, all that
is needed from the noise is the diffusion matrix $D$. Hence, it is
not necessary to require the dimension of the noise vector $\xi$
to be the same as that of the state vector ${\bf q}$. This implies
that in general $l \neq n$. The difficulty for finding such
potential function can be illustrated by the fact that usually
$D^{-1}({\bf q}) {\bf f}({\bf q})$ cannot be written as a gradient
of scalar function \cite{prigogine,vankampen} when no detailed
balance condition is assumed in Eq.(1). Here and below, without
lost of generality the functions, such as ${\bf f}({\bf q})$ and
$D({\bf q})$, are assumed to be sufficiently smooth. The boundary
conditions will be chosen accordingly. This means that boundary
conditions such as absorbing type will not be considered here,
though they can be treated as appropriate limits of the smooth
functions.

%
%

During the study of the robustness of the genetic switch in a
living organism \cite{zhu}, it was discovered that Eq.(1) can be
transformed into the following form,
\begin{equation} \label{normal}
  [ S({\bf q}) + T({\bf q})] \dot{{\bf q}} = - \nabla_{\bf q}
   \phi({\bf q}) + N_{II}({\bf q})\xi(t) \; ,
\end{equation}
where the noise $\xi$ is from the same source as that in Eq.(1).
The $n\times n$ matrices are the symmetric non-negative friction
matrix $S$ and the antisymmetric matrix $T$, and
\begin{equation}
  S({\bf q}) + T({\bf q})
  = \frac{1}{ [ D({\bf q}) + Q({\bf q})] } \equiv M({\bf q}) \; .
\end{equation}
Here $Q$ is an antisymmetric matrix determined by both the
diffusion matrix $D({\bf q})$ and the deterministic force ${\bf
f}({\bf q})$ \cite{kat,ao}. The potential function $\phi({\bf q})$
is connected to the deterministic force ${\bf f}({\bf q})$ by
\begin{equation}
  \nabla_{\bf q} \phi({\bf q}) = - M({\bf q}) {\bf f}({\bf q}) \; .
\end{equation}

The friction matrix $S({\bf q})$ is defined through the following
matrix equation
\begin{equation}
  N_{II}({\bf q}) N_{II}^\tau ({\bf q})= 2 \epsilon \; S({\bf q}) \; ,
\end{equation}
which guarantees that $S$ is both symmetric and nonnegative. For
simplicity we will assume $\det(S) \neq 0$ in the rest of the
paper. It is a sufficient condition for $\det(M) \neq 0$ and more
general cases are also known \cite{kat}. The breakdown of the
detailed balance condition or the time reversal symmetry is
represented by the finiteness of the transverse matrix $T$. The
usefulness of this formulation is already manifested in the
successful solution of outstanding stability puzzle along with new
predictions in gene regulatory dynamics \cite{zhu}.

It was heuristically argued by one of us \cite{ao} that the global
steady-state distribution $\rho({\bf q})$ in the state space is,
if it exists,
\begin{equation} \label{b-g}
 \rho({\bf q}) \propto  \exp\left( - {\phi({\bf q})
   \over{\epsilon} } \right) \; .
\end{equation}
By construction the fixed points of the deterministic force ${\bf
f}$ in Eq.(1) are also the extremal points of the potential
function $\phi$ in Eq.(\ref{normal}) and (\ref{b-g}). Therefore,
the potential function $\phi$ acquires both the local dynamical
meaning through Eq.(\ref{normal}) and the global steady-state
meaning through Eq.(\ref{b-g}). This heuristical demonstration has
been rigorously shown to be locally valid for any fixed point,
stable or unstable \cite{kat}. Two major questions, however,
remain unanswered: Can the heuristical argument be translated into
an explicit procedure such that there is an explicit Fokker-Planck
equation whose steady state solution is indeed given by
Eq.(\ref{b-g})? Is the converse also true, that is, for a given
Fokker-Planck equation, can the corresponding Eq.(\ref{normal}) be
found? Furthermore, are there new and significant results? In this
paper we give affirmative answers to all those important
questions: The general stage is set in Section II; The answer to
the first question is given in section III; The awswer to the
converse question is given in Section IV; and new and significant
results are discussed in Sections III-V.

\section{ Derivation of a generalized Klein-Kramers equation}

Central in the heuristical argument is the introduction of  an
$n$-dimensional kinetic momentum ${\bf p}$ along with a mass $m$.
This procedure brings the stochastic differential equations in
close contact with the Hamiltonian or symplectic structure central
in theoretical physics.  The mass would eventually be taken to be
zero to recover Eq.(\ref{normal}). The dynamical equation for the
enlarged state space is now $2n$ dimensions and the extended
stochastic dynamical equation takes the form \cite{ao}
\begin{eqnarray} \label{langevin}
  \dot{{\bf q}} &=& {{\bf p} \over m} \; , \nonumber \\
  \dot{{\bf p}} &=& -M ({\bf q}){{\bf p} \over m}
  - \nabla_{\bf q} \phi({\bf q}) + N_{II}({\bf q})\xi(t) \; ,
\end{eqnarray}
which is in the form of the standard Langevin physics in the
$({\bf p},{\bf q})$ phase space.
A similar equation has been extensively studied in literature
\cite{vankampen,gardiner}. Here, we investigate it from a
different perspective, the zero-mass limit.

To proceed, we first give an independent derivation of the
generalized Fokker-Planck equation, the so-called Klein-Kramers
equation \cite{vankampen} in a general form, corresponding to
Eq.(\ref{langevin}). We will show that there is no ambiguity in
the treatments of stochastic differential equation at this stage.
The probability distribution function in the $({\bf p},{\bf q})$
phase space is defined by
\begin{equation}
 \rho({\bf p},{\bf q},t)
   \equiv \langle \delta({\bf p}-\bar{\bf p}(t,\{\xi\}))
    \delta({\bf q}-\bar{\bf q}(t,\{\xi\})) \rangle \; ,
\end{equation}
where $\bar{\bf q}(t,\{\xi\})$ and $\bar{\bf p}(t,\{\xi\})$ are
the solution of Eq.(\ref{langevin}) for a given noise
configuration $\{\xi\}$. The distribution function $\rho$ is
obtained by averaging over all the noise configurations, which is
an ensemble average.

With variables $(\bar{\bf q}(t), \bar{\bf p}(t) )$ following
Eq.(\ref{langevin}), the time derivative of the distribution
function $\rho$ is given by
\begin{eqnarray}
 \partial_t \rho({\bf p},{\bf q},t)
  & = & \nabla_{\bf p} \cdot \left[M({\bf q}){{\bf p}
   \over m} + \nabla_{{\bf q}} \phi({\bf q})\right] \rho({\bf p},
   {\bf q},t)-\nabla_{\bf q} \cdot {{\bf p} \over m}
     \rho({\bf p},{\bf q},t) \nonumber \\
  &  & -\nabla_{\bf p} \cdot N_{II}({\bf q})\langle \xi(t)
   \delta({\bf q}-\bar{\bf q}) \delta({\bf p}-\bar{\bf p})
    \rangle \; .  \label{prefp}
\end{eqnarray}
Using an identity due to Novikov \cite{Novikov},
\begin{eqnarray}
 \langle \xi(t) g[\{\xi\}] \rangle
  =  \left\langle {\delta g[\{\xi\}] / \delta \xi(t)} \right\rangle \; ,
\end{eqnarray}
where $g$ is a functional of the noise $\{ \xi \}$, and using the
convention
\begin{equation}
 {\delta \left[\int^t_0 \xi(t') dt \right] / \delta \xi(t)}
  = {1 / 2} \; ,
\end{equation}
and noting that the solution of Eq.(\ref{langevin}) can be
formally expressed as
\begin{eqnarray}
 \bar{\bf q}(t)-{\bf q}(0)
  & = &  \int^t_0 {\bar{\bf p} } dt'/m  \\
  \bar{\bf p}(t)-{\bf p}(0)
  & =  & -\int^t_0 \left[M(\bar{\bf q}){\bar{\bf p} / m}
   - \nabla_{\bar{\bf q}} \phi(\bar{\bf q})
   + N_{II}(\bar{\bf q})\xi \right]dt' ,
\end{eqnarray}
we have the following relations
\begin{eqnarray}
 { \delta \bar{\bf q}(t) / \delta \xi(t) }
   & = & 0 \, , \\
 { \delta \bar{\bf p}(t) / \delta \xi(t) }
   & = & N_{II}^{\tau} (\bar{\bf q})/2 \, .
\end{eqnarray}
The last term in right hand side of equation (\ref{prefp}) is thus
given by
\begin{eqnarray}
 -\nabla_{\bf p} \cdot N_{II}({\bf q})\langle \xi(t)
  \delta({\bf q}-\bar{\bf q}) \delta({\bf p}-\bar{\bf p}) \rangle
 & = &\nabla_{\bf p} \cdot N_{II}({\bf q}) {1 \over 2} N_{II}^{\tau}({\bf q})
  \nabla_{\bf p} \rho({\bf p},{\bf q},t), \label{last-term}
\end{eqnarray}

Combining Eq.(\ref{prefp}) and (\ref{last-term}), we obtain the
Klein-Kramers equation, a special form of the Fokker-Planck
equation,
\begin{eqnarray}
 \partial_t \rho({\bf p},{\bf q},t)
  & = & \nabla_{\bf p} \cdot \left[M({\bf q}){{\bf p} \over m}
   + \nabla_{{\bf q}} \phi({\bf q})+ \epsilon S({\bf q}) \nabla_{\bf p}\right]
   \rho({\bf p},{\bf q},t)-\nabla_{\bf q} \cdot {{\bf p} \over m}
   \rho({\bf p},{\bf q},t) \; . \label{kk-eq}
\end{eqnarray}
A special case of Eq.(\ref{kk-eq}) has been known
\cite{vankampen}. Here we have generalized it to any allowed
matrix $M$. It has the stationary solution, if it exists,
\begin{equation}
 \rho({\bf p},{\bf q})
  = \exp\left(-\frac{[\frac{p^2 }{2m} + \phi({\bf q})] }{ {\epsilon}} \right) \; ,
\end{equation}
which holds for all possible values of mass $m$.

We should point out that starting from Eq.(\ref{langevin}) same
Eq.(\ref{kk-eq}) can be arrived by either Ito or Stratonovich
prescription of stochastic integration, because $\nabla_{\bf
p}^{\tau} M({\bf q}) = 0$. Eq.(19) has been used in the heuristic
demonstration \cite{ao}, to make use of its insensitivity to
various treatments of stochastic differential equation.

\section{ Zero-mass limit and the desired Fokker-Planck equation}

Now we are ready to take the zero-mass limit and to derive the
Fokker-Planck equation corresponding to Eq.(\ref{normal}). We
first define following two operators:
\begin{eqnarray}
 L_1  & \equiv &
   \nabla_{\bf p}^{\tau} M({\bf q}) \left[ \epsilon \nabla_{\bf p}
          + {{\bf p} \over m} \right] \; , \\
 L_2  & \equiv &
   -  {{\bf p} \over m} \cdot \nabla_{\bf q}
   + \nabla_{{\bf q}} \phi({\bf q}) \cdot \nabla_{\bf p} \; .
\end{eqnarray}
With those two operators, Eq.(\ref{kk-eq}) becomes
\begin{equation} \label{rkk-eq}
 \partial_t \rho({\bf p},{\bf q},t)
   =  (L_1 + L_2) \rho({\bf p},{\bf q},t) \; .
\end{equation}
The antisymmetric properties
$
  \nabla_{\bf p}^{\tau} T({\bf q}) \nabla_{\bf p} = 0
$
and
$
 {\bf p}^{\tau} T({\bf q}) {\bf p} = 0
$
are used in above equation.

There are various ways to eliminate the fast degrees of freedom of
${\bf q}$ implied in the zero-mass limit, such as the dynamical
renormalization method \cite{goldenfeld} and the projection
operator method \cite{gardiner,zwanzig,kubo}. In the following, we
adopt from Gardiner \cite{gardiner} the standard projection
operator method for its conciseness. For further exposition of
this method, we refer readers to Ref.\cite{zwanzig,kubo}.

Following Gardiner, we introduce a projection operator
\begin{equation}
 P h({\bf p},{\bf q},t)
  \equiv {1 \over (2\pi m\epsilon)^{n/2}}
   \exp\left(-{p^2 \over {2m\epsilon}} \right)
   \int h({\bf p}',{\bf q},t) d^n p' \; ,
\end{equation}
where $h$ is an arbitrary function of ${\bf p,q}$.

The eigenvalues of the projection operator can only be zero or one,
\begin{equation}
 P^2=P,
\end{equation}
which follows from the relation
\begin{eqnarray}
 P^2 h({\bf p},{\bf q},t) &=& {1 \over (2\pi m\epsilon)^{n/2}}
  \exp\left(-{p^2 \over {2m\epsilon}} \right) \int  {d^n p_1 \over
  (2\pi m\epsilon)^{n/2}} \exp\left(-{p_1^2 \over {2m\epsilon}}
  \right)  \int h({\bf p}_2,{\bf q},t) d^n p_2 \nonumber \\
 & = & {1 \over (2\pi m\epsilon)^{n/2}}
   \exp\left(-{p^2 \over {2m\epsilon}} \right)
   \int h({\bf p}',{\bf q},t) d^n p' \nonumber \\
 & = & P h({\bf p},{\bf q},t).
\end{eqnarray}
From the fact
\begin{equation}
 \left( \epsilon \nabla_{\bf p} + {{\bf p} \over m} \right)
  \exp\left(-{p^2 \over {2m\epsilon}} \right) = 0  \, ,
\end{equation}
we obtain the identity
\begin{equation}
 L_1 P = 0.
\end{equation}
Since $L_1$ is a total derivative operator, for any function $h({\bf
p},{\bf q},t)$ that is well behaved at the boundary (infinity), the
function $P L_1 h({\bf p},{\bf q},t)$ vanishes, because
\begin{eqnarray}
 P L_1 h({\bf p},{\bf q},t) &=& {1 \over (2\pi m\epsilon)^{n/2}}
   \exp\left(-{p^2 \over {2m\epsilon}} \right)
   \int \nabla_{{\bf p}'}^{\tau} M({\bf q}) \left[ \epsilon \nabla_{{\bf
   p}'}+ {{\bf p}' \over m} \right] h({\bf p}',{\bf q},t) d^n p' \nonumber \\
 & = & {1 \over (2\pi m\epsilon)^{n/2}} \exp\left(-{p^2 \over
 {2m\epsilon}} \right) \oint_{\rm B. C.} d {\bf S} \cdot \nabla_{{\bf
  p}'}^{\tau} M({\bf q}) \left[ \epsilon \nabla_{{\bf p}'}
          + {{\bf p}' \over m} \right] h({\bf p}',{\bf q},t)  \nonumber \\
 & = & 0 \nonumber,
\end{eqnarray}
where $d {\bf S}$ is the surface element with direction. From the
last two identities, we can see that the operator $L_1$ is
orthogonal to the projection operator $P$. We further have $
  PL_2P = 0 \; ,
$ due to the inversion symmetry in the ${\bf p}$-space,
\begin{equation}
 \int {d^n p \over (2\pi m\epsilon)^{n/2}}[-  {{\bf p} \over m} \cdot
 \nabla_{\bf q}
   + \nabla_{{\bf q}} \phi({\bf q}) \cdot \nabla_{\bf p}]
   \exp\left(-{p^2 \over {2m\epsilon}} \right)=0.
\end{equation}

To proceed, we first separate the distribution function into the
projected part $v({\bf p},{\bf q},t) \equiv  P\rho({\bf p},{\bf
q},t) $ and unprojected part $ w({\bf p},{\bf q},t) \equiv (1-P)
\rho({\bf p},{\bf q},t) $. We further define the reduced
distribution function $\rho({\bf q},t)$ through the projected
part:
\begin{eqnarray} \label{v-p}
  v({\bf p},{\bf q},t)
   & \equiv & {1 \over (2\pi m\epsilon)^{n/2}}
   \exp\left(-{p^2 \over {2m\epsilon}} \right)
   \rho({\bf q},t) \; .
\end{eqnarray}
The dynamical equations for $v$ and $w$ can be obtained separately
from Eq.(\ref{rkk-eq})
\begin{eqnarray}
 {\partial_t v }
  & = & P\partial_t\rho=P(L_1 + L_2)(v+w)=P L_2 w \; , \label{veq}\\
 {\partial_t w }
  & = & \partial_t\rho-P\partial_t\rho=(L_1+L_2)w + L_2 v - P L_2 w \; .  \label{weq}
\end{eqnarray}
After the Laplace transformation
$
  \tilde{h}(s) = \int_0^\infty h(t) \exp(-st) dt \; ,
$
these two equations take the form
$
 s \tilde{v}-v(0)
   =  P L_2 \tilde{w} $
   and
$ s \tilde{w}-w(0)
   =  (L_1+L_2) \tilde{w}+L_2 \tilde{v}-P L_2
\tilde{w}$.
The latter expression is equivalent to
\begin{equation} \label{twp}
 \tilde{w}
  = [s-L_1-(1-P) L_2]^{-1}[L_2 \tilde{v}+w(0)] \; .
\end{equation}

We note that following Eq.(\ref{langevin}) the relaxation time for
${\bf p}$ dynamics is of the order of $m$. In the zero-mass limit,
this relaxation time is very short. After sufficiently long time,
that is, $t >> m$, which is still short comparing to the dynamics
of the ${\bf q}$, the momentum distribution is essentially
described by the white noise and its fluctuation range is order of
$\sqrt{m}$. Its mean distribution would be determined by the slow
dynamics of ${\bf q}$. Therefore we are looking for the low
frequency behavior of the transformed equation: the leading
contribution when $s << 1/m $. At low frequency, to the leading
order of $m$, the momentum ${\bf p}$ scales with $\sqrt{m}$, $L_1$
is of the order of $1/m$, and $L_2$ is of the order of
$1/\sqrt{m}$.  Hence, at low frequency to the leading contribution
ordered by $m$, Eq.(\ref{twp}) leads to
\begin{equation}
 \tilde{w}= -L_1^{-1}L_2 \tilde{v} + O(m) \; ,
\end{equation}
which is a precise statement on the adiabatic following of kinetic
momentum ${\bf p}$ to the coordinate ${\bf q}$. The equation for
$v$ is thus given by
\begin{equation} \label{eqv}
 {\partial_t v } = -P L_2 L_1^{-1} L_2 v + O(\sqrt{m}) \; .
\end{equation}

We recall an identity to be used. The operator $L_1$ has a null
space and its inverse operator is not well defined unless in the
space orthogonal to the null space. For an arbitrary vector ${\bf
c}({\bf q})$ which has no ${\bf p}$-dependence, the following
identity holds
\begin{eqnarray} \label{L1-eq}
 L_1 {\bf p} \cdot {\bf c}({\bf q}) \exp\left(-{p^2 \over {2m\epsilon} }\right)
  & = & \epsilon \nabla_{\bf p} \cdot M({\bf q}) {\bf c}({\bf q})
   \exp\left(-{p^2 \over {2m \epsilon} }\right)  \nonumber \\
  & = & -{{\bf p} \over {m } } \cdot M({\bf q}) {\bf c}({\bf q})
   \exp\left(-{p^2 \over {2m\epsilon}}\right).
\end{eqnarray}
We note that $L_2 v$ takes the form of the right hand side of
Eq.(\ref{L1-eq}), and is therefore orthogonal to the null space of
$L_1$. The inverse operator $L_1^{-1}$ is then well defined. Using
the inverse relation of Eq.(\ref{L1-eq}) we arrive at the desired
identity:
\begin{equation}
 L_1^{-1} {{\bf p} \over {m } } \cdot {\bf c}({\bf q})
   \exp\left(-{p^2 \over {2m \epsilon} }\right)
  = -{\bf p} \cdot M^{-1}({\bf q}){\bf c}({\bf q})
   \exp\left(-{p^2 \over {2m\epsilon} }\right) \; .
\end{equation}

With above identity, the right hand side of Eq.(\ref{eqv}) is
given by
\begin{eqnarray}  \label{L1}
 -P L_2 L_1^{-1} L_2 v
  & = & P L_2 L_1^{-1} {{\bf p} \over m} \cdot
   [\nabla_{\bf q}+ {1 \over \epsilon }\nabla_{{\bf q}} \phi({\bf q})] v \;
   \nonumber \\
  & = & \nabla_{{\bf q}} \cdot M^{-1}({\bf q}) [\epsilon \nabla_{\bf q}
   + \nabla_{{\bf q}} \phi({\bf q})] v \; .
\end{eqnarray}

Therefore in the zero-mass limit, $ m \rightarrow 0 $, the
equation for the integrated probability distribution $\rho({\bf
q},t)$ defined in Eq.(\ref{v-p}) takes the form, as a direct
consequence of Eq.(\ref{eqv}) and (\ref{L1})
\begin{equation} \label{fp-eq}
 {\partial_t \rho({\bf q},t) }
  = \nabla_{{\bf q}} M^{-1}({\bf q}) [\epsilon \nabla_{\bf q}
   + \nabla_{{\bf q}} \phi({\bf q})] \rho({\bf q},t) \; .
\end{equation}
This is the sought Fokker-Planck equation corresponding to
Eq.(\ref{normal}).
We point out that in the above derivation we take the mass to be
zero, keeping other parameters, including the friction and
transverse matrices, finite. On the other hand, in the usual
Smoluchowski limit it is the friction matrix that has to be taken
as infinite, keep all other parameters finite. Those two limits
are in general not interchangeable.

The equilibrium configuration solution of Eq.(\ref{fp-eq}) is the
same as Eq.(\ref{b-g}).
Again, we emphasize that no detailed balance condition is assumed
in reaching this result. This completes our answer to the first
question of finding the corresponding Fokker-Planck equation.

\section{Converse problem}

We now address the second main question that for any given
Fokker-Planck equation there is the corresponding stochastic
differential equation, Eq.(\ref{normal}). We will give an
affirmative answer, which closes a logic gap in the light of
present formulation. The procedure to carry it out is already
implicitly contained in Eq.(\ref{fp-eq}), a typical situation in
theoretical physics that if the answer is known a procedure to
obtain it can be easily found. In addition, the demonstration in
this section also supplements above rather abstract projection
operator demonstration.

A generic Fokker-Planck equation for the dynamics of probability
density in state space may take the form:
\begin{equation} \label{fp-g}
  {\partial_t \rho({\bf q},t) }
  = \nabla_{{\bf q}}^{\tau}
   [\epsilon  \overline{D}({\bf q}) \nabla_{\bf q}
    - \overline{{\bf f}}({\bf q})] \rho({\bf q},t) \; .
\end{equation}
Here $\overline{D}({\bf q})$ is the diffusion matrix and
$\overline{{\bf f}}({\bf q})$ the drift force. A potential
function $\overline{\phi}({\bf q})$ can always be defined from the
steady state distribution. This has been extensively studying in
mathematics \cite{doob}. Given the existence of the potential
function, the procedure is particularly simple.

Using $ M^{-1}({\bf q})= D({\bf q}) + Q({\bf q})$ \cite{ao},
Eq.(\ref{fp-eq}) can be rewritten as
\begin{equation}  \label{fp-eq2}
 {\partial_t \rho({\bf q},t) }
  = \nabla_{{\bf q}}^{\tau} [\epsilon D({\bf q})\nabla_{\bf q}
   - \epsilon (\nabla_{{\bf q}}^{\tau} Q({\bf q}))^{\tau}
   +  [D({\bf q}) + Q({\bf q})]\nabla_{{\bf q}} \phi({\bf q})]
     \rho({\bf q},t) \; .
\end{equation}
The antisymmetric property of the matrix $Q({\bf q})$ has been
used in reaching Eq.({\ref{fp-eq2}).  Thus, comparing between
Eq.(\ref{fp-g}) and (\ref{fp-eq2}), we have $ D({\bf q})
  = \overline{D}({\bf q}) \; ,
\phi({\bf q}) = \overline{\phi}({\bf q}) \; , $ and
\begin{eqnarray}
  {\bf f}({{\bf q}})
   & = & \overline{{\bf f}}({\bf q})
    - \epsilon (\nabla_{{\bf q}}^{\tau} Q({\bf q}))^{\tau} \; . \label{force}
\end{eqnarray}
In reaching Eq.(\ref{force}) we have used the relation $ [D({\bf
q}) + Q({\bf q})] \nabla_{{\bf q}} \phi({\bf q}) = - {\bf f}({\bf
q}) $.  The explicit equation for the anti-symmetric matrix $Q$ is
\begin{equation} \label{Q-eq}
 - \epsilon (\nabla_{{\bf q}}^{\tau} Q({\bf q}))^{\tau}
 + [ D({\bf q}) + Q({\bf q}) ]  \nabla_{{\bf q}} \phi({\bf q})
    = - \overline{{\bf f}}({\bf q}) \; .
\end{equation}
The solution for $Q$ can be formally written down
\begin{equation}
  Q({\bf q}) = - {1 \over{\epsilon}} \int^{\bf q} d{\bf q}'
   [ \overline{\bf f}({\bf q}')
    + D({\bf q}')\nabla_{{\bf q}'} \phi({\bf q}') ]
    \exp\left( {\phi({\bf q}')  - \phi({\bf q}) \over{\epsilon}} \right)
  + Q_0({\bf q}) \exp\left(- { \phi({\bf q}) \over{\epsilon}} \right) \; .
\end{equation}
Here $Q_0({\bf q})$ is a solution of the homogenous equation
$\epsilon \nabla_{{\bf q}}^{\tau} Q({\bf q}) = 0 $ and the two
parallel vectors in the integrand, such as $d{\bf q}' \;
\overline{\bf f}({\bf q})$, forms a matrix. This completes our
answer to the converse question of finding the corresponding
stochastic differential equation in the form of Eq.(\ref{normal})
from any given Fokker-Planck equation.

We note that the shift between the zero's of the potential
gradient and the drift force is given by, from Eq.(\ref{force}),
\begin{equation} \label{shift}
 \Delta \overline{\bf f}
  = - \epsilon (\nabla_{{\bf q}}^{\tau} Q({\bf q}))^{\tau} \; ,
\end{equation}
that is, the extremals of the steady state distribution are not
necessarily determined by the zero's of drift force $
\overline{\bf f} $. To our knowledge this is the first time that
such an analytic formulae for the shift is obtained.

It is worthwhile to point out that a construction similar to that
of above was discussed in Ref. \cite{eyink}. In order to obtain
the desired potential function, several additional conditions,
including one similar to set $\nabla_{{\bf q}} Q({\bf q}) =0 $
(their (4.18)), were required in Ref. \cite{eyink}. Our present
demonstration shows that there is no need for those conditions.
Hence, our construction may be regarded as a generalization of the
corresponding one in Ref. \cite{eyink}.

\section{ Discussions}

Attempts to decompose the dynamics into the dissipative and
transverse parts were extensively explored in literature in the
framework of Fokker-Planck equation \cite{go,pp}. Though
conceptually the basic ideas in literature are similar to what
discussed here, the present demonstration shows that in general
there is no apparent separation between the friction and the
transverse matrices implied in those previous works, because the
gradient of the antisymmetric matrix $Q$ in Eq.(41) is in general
not zero. The anti-symmetric matrix $Q$ should be determined by
both diffusion matrix $D$ and deterministic force $\bf f$ in
Eq.(1), or, by both friction and transverse matrices in
Eq.(\ref{normal}). Furthermore, the connection between the local
micro-dynamics describing by Eq.(\ref{normal}) and the global
macro-dynamics discussed in Eq.(41), or Eq.(42) or (43), was not
discussed in Ref.\cite{go,pp}. In fact, the present authors did
not aware such a connection prior to 2004 \cite{kat,ao}. We should
remark here that the special form of the stochastic differential
equation, Eq.(\ref{normal}), is consistent with the formulation of
dissipative dynamics from first principles \cite{zwanzig,leggett}.



If the antisymmetric matrix $Q$ is zero, there would be no shift
between the zero's of drift force and the potential gradient
according to Eq.(\ref{fp-eq}) and (\ref{fp-g}). The drift force in
this case can be expressed as $ \overline{{\bf f}}({\bf q}) = -
D({\bf q}) \nabla_{\bf q} \phi({\bf q})$, exactly the detailed
balance condition. However, even if $D$ is independent of the
state vector, that is, there is no difference between Ito and
Stratonovich treatments of stochastic differential equations, the
anti-symmetric matrix $Q$ can still be state vector dependent.
There would still be a shift between the zero of the potential
gradient and the drift force. This is precisely what have been
found in numerical studies on noise induced phase transitions and
bifurcations \cite{shift}. Eq.(\ref{shift}) is a formula for this
shift, which appears for the first time in the present letter.

There is an apparent disagreement between the singular behaviors
found in the escape path study \cite{dykman,ban} and a possible
smooth potential function implied in the present study. While a
detailed study on this feature is beyond the present letter, which
will be reported elsewhere, we point out two main factors which
are responsible for this apparent disagreement. The first factor
is the difference in specifying the stochastic integration
procedures. This difference results in a shift between the zeros
of drift force and extremals of the steady steady distribution,
described by the shift formula, Eq.(\ref{shift}). The second
factor is that in Ref. \cite{dykman} and \cite{ban} the focus is
on the escaping rate and the corresponding escaping path, not on
the steady state distribution. The emergence of singularity is
then not surprising, because its sensitivity to the dynamical
elements, the transverse matrix $T$ and the friction matrix $S$,
in additional to the noise strength specified by $\epsilon$.

Finally, there is another immediate and testable prediction from
the present formulation. The limit cycle dynamics, abundant in
nonequilibrium processes, has been used as a prototype example to
argue against the existence of potential function. Not only our
formulation suggests its existence in the sense of
Eq.(\ref{normal}), (\ref{b-g}), and (\ref{fp-eq}), which is
natural in theoretical physics, also it should take the same value
along the limit cycle \cite{ao05}.

{\ }

We thank critical and constructive discussions with D.J. Thouless,
X.-M. Zhu, H. Qian, C. Kwon, M. Dykman, Q. Ouyang, R. Mannella and
P.V.E. McClintock during the course of our investigation. We also
thank G.L. Eyink for calling our attention to Ref. \cite{eyink}
and for a following-up clarifying discussion. This work was
supported in part by USA NIH grant under HG002894 and by China
NSFC under grant number 90303008.

\end{document}